\documentclass{webofc}
\usepackage[varg]{txfonts}   
\newcommand{\sqrts}{\sqrt{s}}
\newcommand{\sqrtsNN}{\sqrt{s_{\rm \scriptscriptstyle NN}}}

\newcommand{\gevc}{\mathrm{GeV}/c}

\newcommand{\Raa}{R_{\rm AA}}

\newcommand{\pt}{p_{\rm T}}

\newcommand{\Dzero}{{\rm D^0}}
\newcommand{\Dstar}{{\rm D^{*+}}}
\newcommand{\Dplus}{{\rm D^+}}
\newcommand{\Lc}{\Lambda^+_{\rm c}}
\newcommand{\Ds}{{\rm D}^+_{\rm s}}

\begin{document}
\title{Open heavy-flavour measurements with ALICE at the LHC}
%
%

\author{\firstname{First author} \lastname{Fabio Colamaria}\inst{1}\fnsep\thanks{\email{fabio.colamaria@ba.infn.it}} on behalf of the ALICE Collaboration
}

\institute{Istituto Nazionale di Fisica Nucleare, Sezione di Bari - Via E. Orabona 4, 70125 Bari, Italy}

\abstract{%
Heavy quarks are produced in the early stages of ultra-relativistic heavy-ion collisions, and their number is preserved throughout the subsequent evolution of the system. Therefore, they constitute ideal probes for characterising the Quark--Gluon Plasma (QGP) medium and for the study of its transport properties. In particular, heavy quarks interact with the partonic constituents of the plasma, losing energy, and are expected to be sensitive to the medium collective motion induced by its hydrodynamical evolution.
In pp collisions, the measurement of heavy-flavour hadron production provides a reference for heavy-ion studies, and allows also testing perturbative QCD calculations in a wide range of collision energies.
Similar studies in p--Pb collisions help in disentangling cold nuclear matter effects from modifications induced by the presence of a QGP medium, and are also useful to investigate the possible existence of collective phenomena also in this system.
The ALICE detector provides excellent performances in terms of particle identification and vertexing capabilities. Hence, it is fully suited for the reconstruction of charmed mesons and baryons and of electrons from heavy-flavour hadron decays at central rapidity. Furthermore, the ALICE muon spectrometer allows reconstructing heavy-flavour decay muons at forward rapidity.
A review of the main ALICE results on open heavy flavour production in pp, p--Pb and Pb--Pb collisions is presented. Recent, more differential measurements are also shown, including azimuthal correlations of heavy-flavour particles with charged hadrons in p--Pb collisions, and D-meson tagged-jet production in p--Pb and Pb--Pb collisions.
}
\maketitle
\section{Introduction}
\label{intro}
Heavy quarks (charm and beauty) are excellent probes for the study of the Quark--Gluon Plasma (QGP), a colour-deconfined medium of strongly-interacting matter formed in ultrarelativistic heavy-ion collisions~\cite{BraunMunzinger:2007zz}.
Due to their large mass, heavy quarks are produced in hard parton scatterings during the initial stages of the collision, and their thermal production and in-medium annihilation rates are negligible at the LHC energies. Hence, they experience the whole evolution of the QGP medium and are subject to its effects.

In particular, charm and beauty quarks are expected to interact with the medium constituents via elastic and inelastic processes. This leads to a loss of their initial energy while crossing the medium~\cite{Gyulassy:1990ye,Baier:1996sk,Thoma:1990fm}, though to a lesser extent with respect to lighter quarks and gluons, due to the colour-charge and `dead-cone' effects~\cite{Armesto:2005iq}.
A sensitive observable to partonic energy loss is the nuclear modification factor, defined as $R_{\rm AA}(\pt) = \frac{{\rm d}N_{\rm AA}/{\rm d}\pt}{\langle N_{\rm coll} \rangle {\rm d}N_{\rm pp}/{\rm d}\pt}$, i.e. the ratio of the production yields of a given particle normalised by the production yield in pp collisions and scaled by the average number of binary nucleon-nucleon collisions $\langle N_{\rm coll} \rangle$.

In heavy-ion collisions with non-zero impact parameter, a collective motion of the medium is also expected to emerge, in case of thermalisation of its constituents. This motion translates the initial spatial anisotropy of the system into a momentum anisotropy, reflected in the final-state particle azimuthal distribution~\cite{Ollitrault:1992bk}. The degree of participation of heavy quarks to the medium collective motion can be quantified via the `elliptic-flow' coefficient $v_2$ at low and intermediate transverse momentum ($\pt$). The elliptic-flow coefficient is defined as the second-order coefficient of the Fourier decomposition of the azimuthal distribution of a given particle.

Among the heavy-flavour species, the study of charm and beauty baryons is of particular interest. It was suggested that, in a QGP environment, heavy-flavour quarks could hadronise also via recombination with other quarks present in the medium, in addition to string fragmentation. This would induce an enhancement of the baryon-to-meson ratio at low and intermediate $\pt$ with respect to pp collisions~\cite{Greco:2003vf,Oh:2009zj}.
In addition, for pp collisions, the sensitivity of heavy-flavour baryons to fragmentation allows us to test the universality of the parton fragmentation functions, commonly evaluated from electron-positron data.

The availability of large data samples collected during the latest years has opened the possibility of performing more differential analysis also for rare probes as heavy flavours.
In this contest, the study of azimuthal correlation distributions between electrons from heavy-flavour hadron decays (HFe) and charged hadrons in p--Pb collisions can provide a measurement of the HFe $v_2$ coefficient also in this collision system. A positive $v_2$ was already measured by ALICE for light-flavour hadrons at midrapidity~\cite{Abelev:2012ola}, and its interpretation, either as a final-state or initial-state effect, is currently debated, evidencing the need of additional measurements also in the heavy-flavour sector.
Another recent differential study is the measurement of the production cross section of jets with a D-meson constituent (D-jets), in all collision systems. This measurement provides a closer access to the parton kinematics with respect to the study of the D-meson cross section. In addition, comparing the momentum fraction of the jet carried by the D meson in pp and Pb--Pb collisions provides information on the modification of the heavy-quark fragmentation in a QGP medium.

\section{Experimental apparatus and results}
\label{sec-main}
The ALICE detector~\cite{Aamodt:2008zz,Abelev:2014ffa} provides excellent performances in terms of particle identification, low-$\pt$ track reconstruction, and vertexing. Thanks to these capabilities, open heavy-flavour particles at low and intermediate $\pt$ can be reconstructed with great precision.
In particular, $\Dzero, \Dstar, \Dplus$ and $\Ds$ mesons, as well as $\Lambda^+_{\rm c}$ and $\Xi^0_{\rm c}$ baryons can be reconstructed from their hadronic (for D-mesons and $\Lambda^+_{\rm c}$ baryons) and semileptonic (for $\Lambda^+_{\rm c}$ and $\Xi^0_{\rm c}$) decay channels at midrapidity ($|y|<0.5$), exploiting their displaced decay topologies, thanks to the possibility of separating primary and secondary vertices with the Inner Tracking System (ITS). To improve the selection, daughter particles can be identified
by means of specific energy loss in the Time Projection Chamber (TPC) and time-of-flight in the Time Of Flight (TOF) detector.
Heavy-flavour hadron decay electrons can also be reconstructed at central rapidity. The identification of electron at high $\pt$ is performed with the Electromagnetic Calorimeter (EMCal), while TPC and TOF are used in the low $\pt$ region.
Furthermore, the ALICE muon spectrometer, composed of a set of tracking and trigger stations, allows reconstructing heavy-flavour hadron decay muons at forward rapidity ($-4 < y < -2.5$).

\subsection{Results for charmed mesons}
\label{sub1}
\begin{figure}[h]
\centering 
\includegraphics[width=0.50\linewidth,clip]{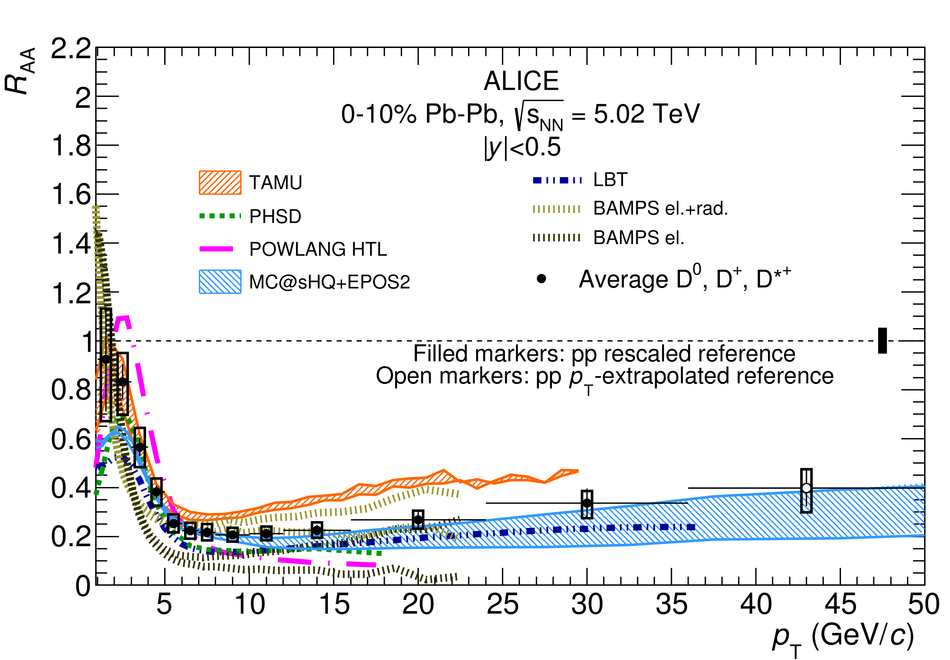}
\includegraphics[width=0.48\linewidth,clip]{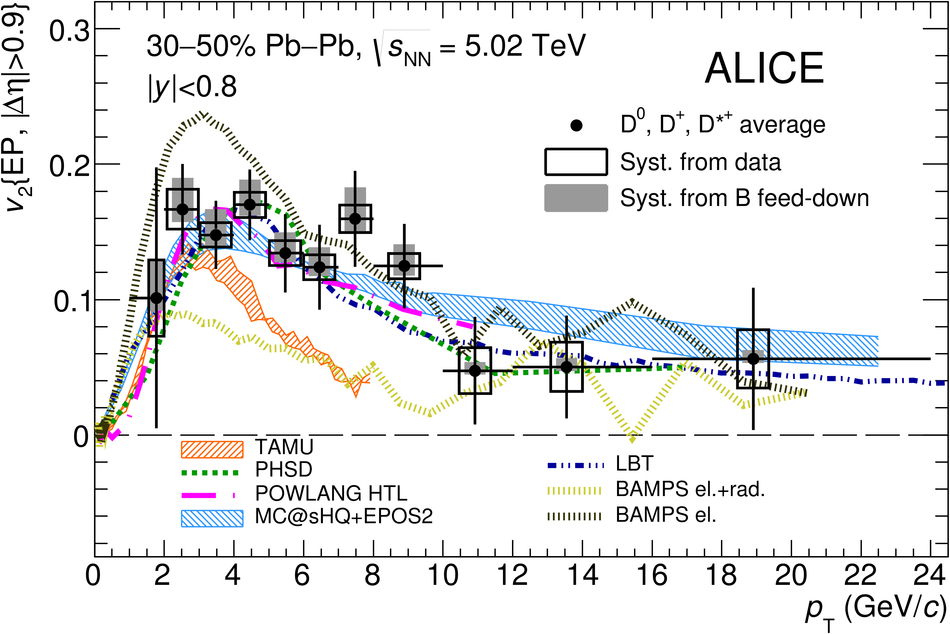}
\caption{Left: average nuclear modification factor of $\Dzero$, $\Dplus$ and $\Dstar$ mesons as a function of $\pt$ in 0--10\% central Pb--Pb collisions. Right: average elliptic-flow coefficient of $\Dzero$, $\Dplus$ and $\Dstar$ mesons in Pb--Pb collisions, for 30--50\% centrality. For both panels, data are compared with model predictions which provide a simultaneous description of both observables.}
\label{Dmeson_Raa}       
\end{figure}

The left panel of Fig.~\ref{Dmeson_Raa} shows the nuclear modification factor for the average of non-strange D mesons as a function of $\pt$, in 0--10\% central Pb--Pb collisions at $\sqrtsNN = 5.02$ TeV~\cite{Acharya:2018hre}. A strong suppression of a factor $\approx5$ is observed for $2 < \pt < 6$ $\gevc$, while the $R_{\rm AA}$ increases at lower $\pt$, being compatible with unity for $1 < \pt < 2$ $\gevc$. This suppression can be ascribed to final-state effects, since in p--Pb collisions the equivalent observable $R_{\rm pPb}$ was found to be compatible with unity within uncertainties over the full $\pt$ range~\cite{Adam:2016ich}.
On the right panel of the same figure, the elliptic-flow coefficient of the average of non-strange D mesons is shown for Pb--Pb collisions at $\sqrtsNN = 5.02$ TeV, in the 30--50\% centrality class. A positive $v_2$ is measured in the range $2 < \pt < 10$ $\gevc$, with a strength comparable to that of charged pions. This suggests that low-momentum charm quarks effectively participate to the medium collective motion.
Both measurements refer to the prompt component of D-meson yields only. The feed-down contribution was subtracted via a model-driven approach, based on FONLL~\cite{Cacciari:1998it} pQCD calculations for feed-down D-meson production.

In both panels, ALICE measurements are compared with model predictions based on heavy-quark transport in a viscous hydrodynamic medium, which provide a description for both observables. Models in which charm quarks pick up collective flow via recombination or subsequent elastic collisions and include radiative energy loss in an expanding medium better reproduce the two observables. Anyway, a simultaneous description of both observables over the full $\pt$ range remains a challenging task, and these measurements are able to set strong constraints to models.
The best $v_2$ description is provided by models predicting a charm thermalisation time of $\tau_{\rm c} = 3 - 14$ ${\rm fm}/c$. This is of the same order of the QGP decoupling time, implying a certain degree of thermalisation of charm quarks in the medium.

\subsection{Results for charmed baryons}
\label{sub2}
\begin{figure*}
\centering
\includegraphics[width=0.57\linewidth,clip]{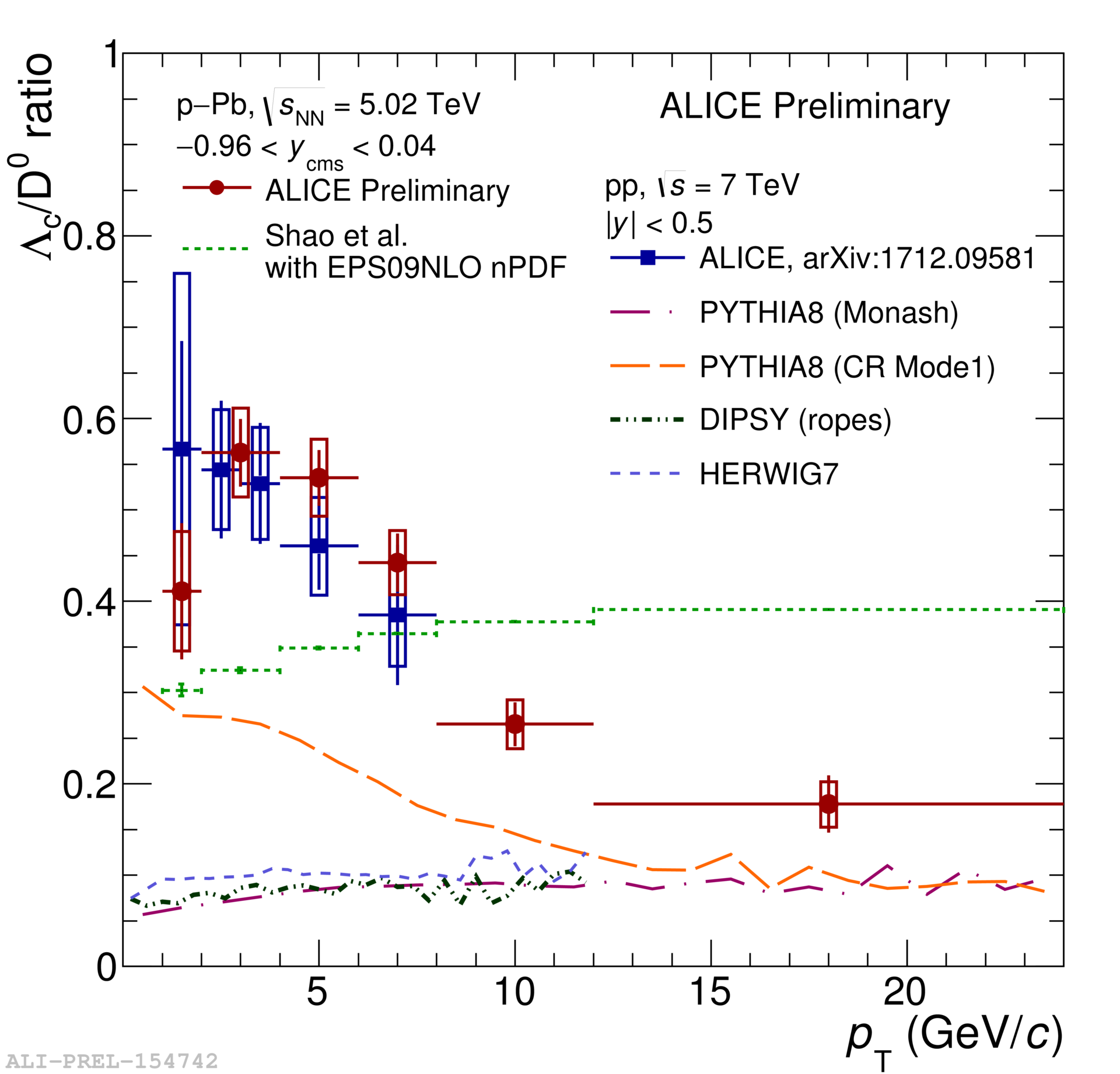}
\caption{Ratio of $\Lambda^+_{\rm c}/ \Dzero$ production yields in pp and p--Pb collisions as a function of $\pt$, compared with Monte Carlo predictions.}
\label{LcDoRatio}       
\end{figure*}

In Fig.~\ref{LcDoRatio} ALICE results for the ratio of $\Lc$ and $\Dzero$ prompt production cross sections are presented as a function of $\pt$, for pp collisions at $\sqrts = 7$ TeV~\cite{Acharya:2017kfy} and p--Pb collisions at $\sqrtsNN = 5.02$ TeV (preliminary results from LHC Run 2). The two measurements are compatible within uncertainties over the full $\pt$ range. They also show the same transverse momentum evolution observed in the light-flavour sector for the $\Lambda/{\rm K^0_{\rm s}}$ ratio~\cite{Abelev:2013xaa,Adam:2016dau}, with a decreasing trend for $\pt > 4$ $\gevc$.
ALICE results show larger ratios with respect to measurements performed in e$^+$e$^-$ collisions at lower centre-of-mass energies, where mechanisms expected to increase the baryon-to-meson ratio as colour reconnection and multi-parton interactions are expected to play a smaller role. The $\Lc/\Dzero$ ratios reported by ALICE are also larger, though still compatible, than analogue measurements by LHCb~\cite{Aaij:2018iyy} at forward rapidity ($2 < y < 4.5$) in a comparable transverse momentum range and at the same energy.

Measurements in pp collisions are compared to predictions from several Monte Carlo generators employing different hadronisation mechanisms, with fragmentation fractions of charm into hadrons based on electron-positron collision data. The event generator which provides the closest description to data is PYTHIA8~\cite{Sjostrand:2006za,Sjostrand:2007gs} with an enhanced colour reconnection mechanism (orange dashed curve), which is supposed to increase the baryon production. Nevertheless, also this generator underestimates ALICE measurements by a factor $\approx2$ over the full $\pt$ range.

\begin{figure*}
\centering
\includegraphics[width=0.97\linewidth,clip]{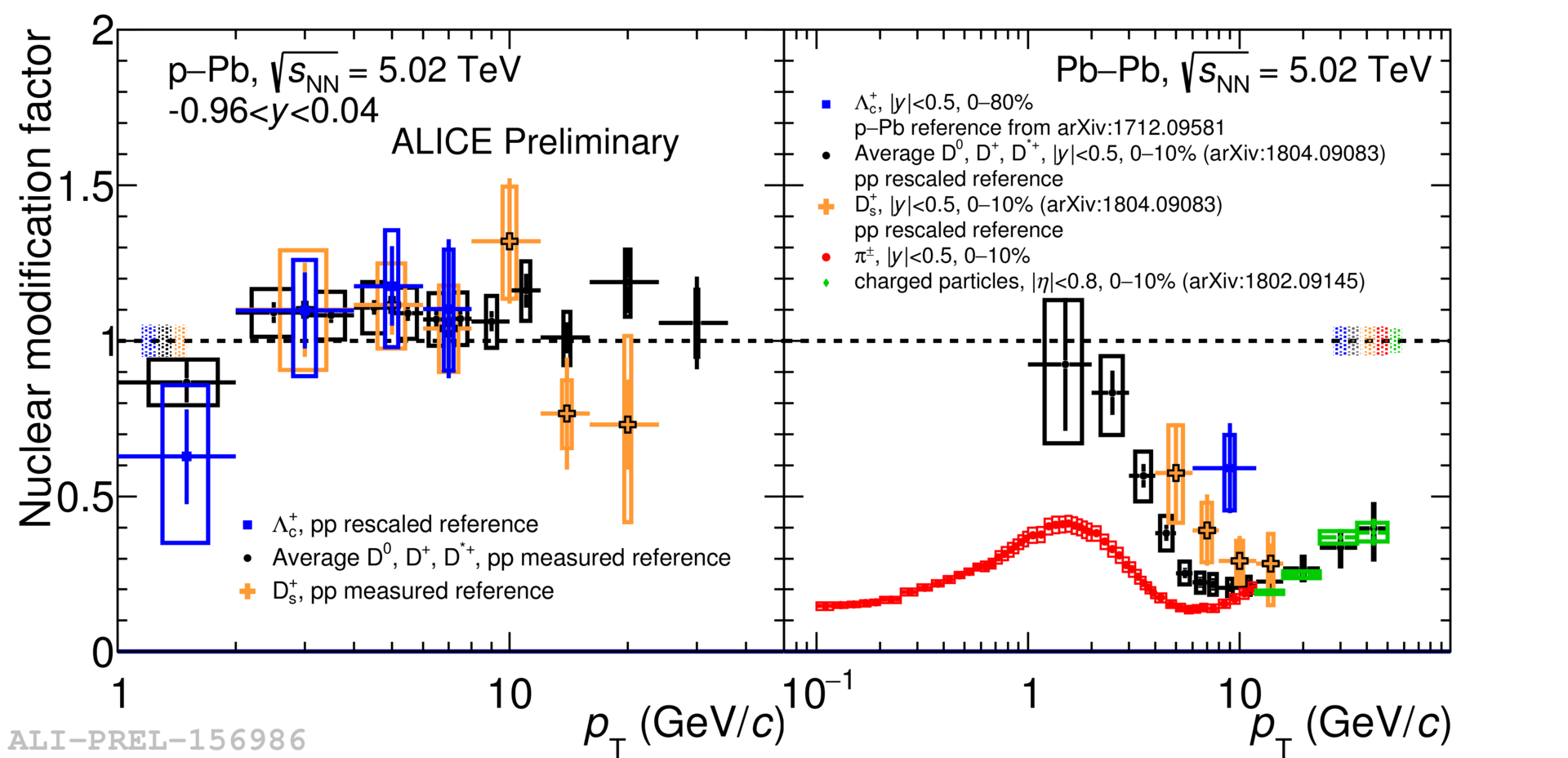}
\caption{Comparison of nuclear modification factor of average D mesons, $\Ds$, $\Lambda^+_{\rm c}$, charged pions and charged particles in p--Pb (left panel) and Pb--Pb (right panel) collisions.}
\label{D_Ds_Lc}       
\end{figure*}

Figure~\ref{D_Ds_Lc} shows ALICE measurements of the nuclear modification factor of non-strange D mesons (average of $\Dzero$, $\Dplus$ and $\Dstar$), $\Ds$ meson and $\Lc$ as a function of $\pt$, at central rapidity, in p--Pb collisions (left panel) and Pb--Pb collisions (right panel) at $\sqrtsNN = 5.02$ TeV. 0--10\% (0--80\%) central Pb--Pb collisions are considered for D-meson ($\Lc$) measurement.
The values of $R_{\rm pPb}$ of all the charmed hadron species are consistent within uncertainties and compatible with unity, suggesting a limited impact of cold-nuclear-matter effects. The measurements in Pb--Pb show a possible hierarchy for the nuclear modification factor values, with $R_{\rm AA}(\Lc) > R_{\rm AA}(\Ds) > R_{\rm AA}$(non-strange D). This feature could be explained by the onset of coalescence mechanism in Pb--Pb central collisions: $\Ds$ production should be increased by recombination of charm quark with strange quarks in a strangeness-rich environment as the QGP, and coalescence shall enhance the production of baryons with respect to the production rate in pp collisions.

Charmed-hadron $\Raa$ is also compared to nuclear modification factor measurements for charged tracks and charged pions. A hint of a larger $\Raa$ is observed for D mesons with respect to charged pions over the common $\pt$ range. It is not straightforward, anyway, to interpret this feature as a smaller in-medium energy loss for charm quarks rather than for light-flavour quarks and gluons, due to the dead-cone effect. The light hadron production does not scale with the average number of binary collision $\langle N_{\rm coll} \rangle$ for $\pt \leq 3 \gevc$; a different energy scale from heavy- and light-flavour hadrons to corresponding partons could enter in play due to the different fragmentation of the latter, and the initial $\pt$ spectrum is not the same for light and heavy partons; finally, the radial flow contribution could affect differently light- and heavy-flavour mesons.

\subsection{Results for heavy-flavour correlations and jets}
\label{sub3}
\begin{figure}[h]
\centering
\includegraphics[width=0.47\linewidth,clip]{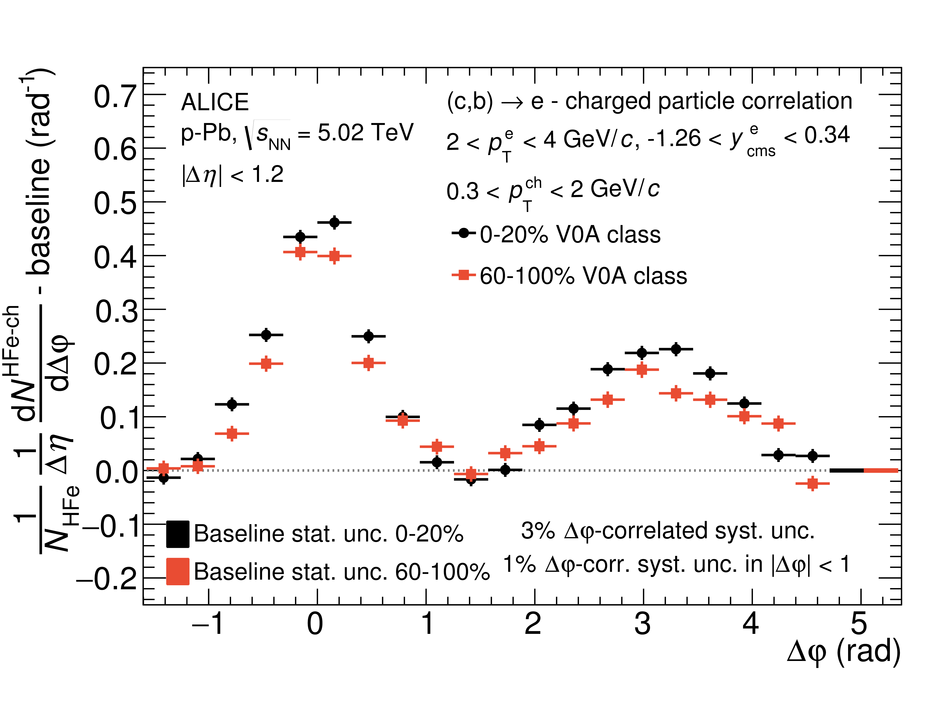}
\includegraphics[width=0.51\linewidth,clip]{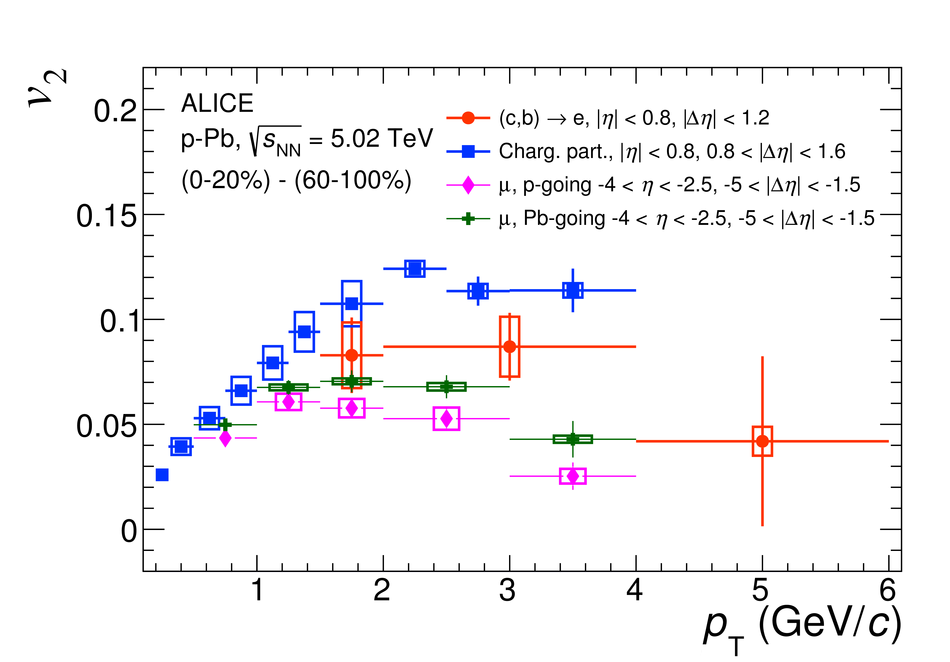}
\caption{Left: baseline-subtracted azimuthal correlations between HFe and charged particles, for high-multiplicity and low-multiplicity p--Pb collisions. Right: $v_2$ of HFe as a function of $\pt$ compared to the $v_2$ of unidentified charged particles.}
\label{Hfe_h}       
\end{figure}

The left panel of Figure~\ref{Hfe_h} shows the azimuthal correlation distribution between heavy-flavour decay electrons with $2 < \pt < 4$ $\gevc$ and charged particles with $0.3 < \pt < 2$ $\gevc$, in p--Pb collisions at $\sqrtsNN = 5.02$ TeV, normalised by the number of heavy-flavour decay electrons~\cite{Acharya:2018dxy}. Two different multiplicity ranges, 0--20\% (red points) and 60--100\% (black points), classified using the signal amplitude in the V0-A detector, were considered. The correlation distributions are shown after the subtraction of a `baseline' term, evaluated via a fit with a constant to the three lowest points of the distribution. An enhancement of the peaks at $\Delta\varphi \approx 0$ and $\Delta\varphi \approx \pi$ can be observed for high-multiplicity collisions. The jet contribution is assumed comparable for the two multiplicity classes and removed by subtracting the low-multiplicity from the high-multiplicity distribution. This allows extracting the $v_2$ coefficient of heavy-flavour decay electrons via a fit to the subtracted correlation distribution with a Fourier decomposition in $\Delta\varphi$.

The values of the elliptic-flow coefficient are shown in the right panel of Fig.~\ref{Hfe_h} as a function of the electron transverse momentum. A positive $v_2$ is measured for the range $1.5 < \pt < 4$ $\gevc$ with a significance of 5.1$\sigma$. The values of heavy-flavour decay electron $v_2$ are generally lower, though still comparable, than those measured for charged particles, dominated by light-flavour hadrons. An interpretation of this comparison is anyway not immediate, since the transverse-momentum distribution of heavy-flavour hadrons from which the electrons originate is considerably broader than that of light-flavour hadrons, as well as having a larger average value.
The $v_2$ of heavy-flavour decay electrons has a similar trend of that of inclusive muons, measured by ALICE at forward and backward rapidity~\cite{Adam:2015bka} and dominated by heavy-flavour decay muons for $\pt > 2$ $\gevc$. Drawing conclusions from this comparison is not straightforward because of the different cold-nuclear-matter effects affecting heavy flavours in different rapidity ranges.

ALICE has measured the $\pt$-differential production cross section of prompt $\Dzero$-jets in p--Pb and Pb--Pb collisions at $\sqrtsNN = 5.02$ TeV. Jets were reconstructed in the pseudorapidity range $|\eta| < 0.6$, using the FASTJET~\cite{Cacciari:2011ma} package with the anti-$k_{\rm T}$ algorithm, after substituting the daughters tracks of the $\Dzero$ with the $\Dzero$ particle. The feed-down contribution to the $\Dzero$-jet spectrum was subtracted exploiting POWHEG+PYTHIA~\cite{Nason:2004rx,Jadach:2015mza} simulations. Detector effects and background fluctuations were corrected for with a Bayes unfolding procedure~\cite{DAgostini:2010hil}.

\begin{figure}[h]
\centering
\includegraphics[width=0.49\linewidth,clip]{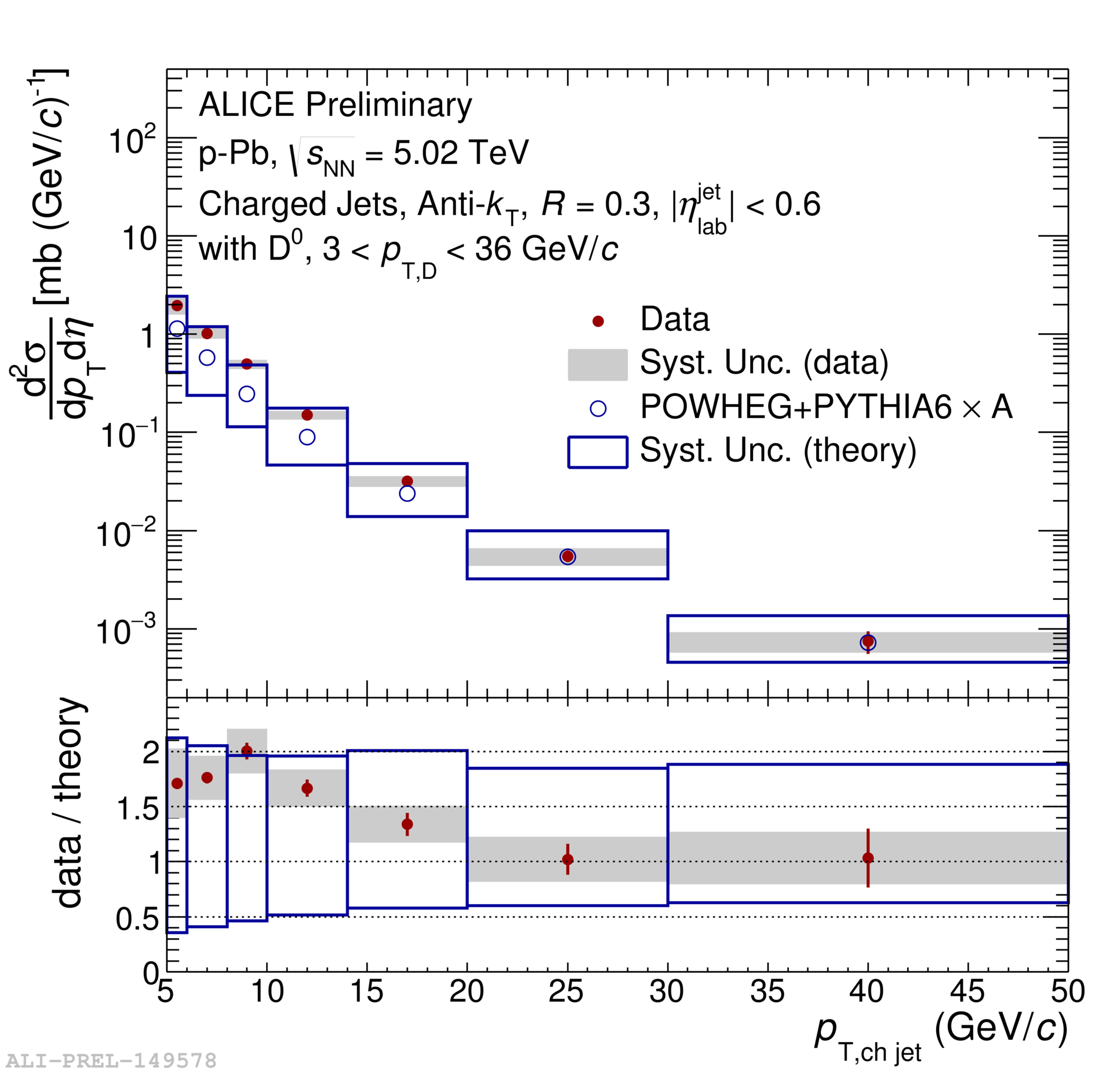}
\includegraphics[width=0.49\linewidth,clip]{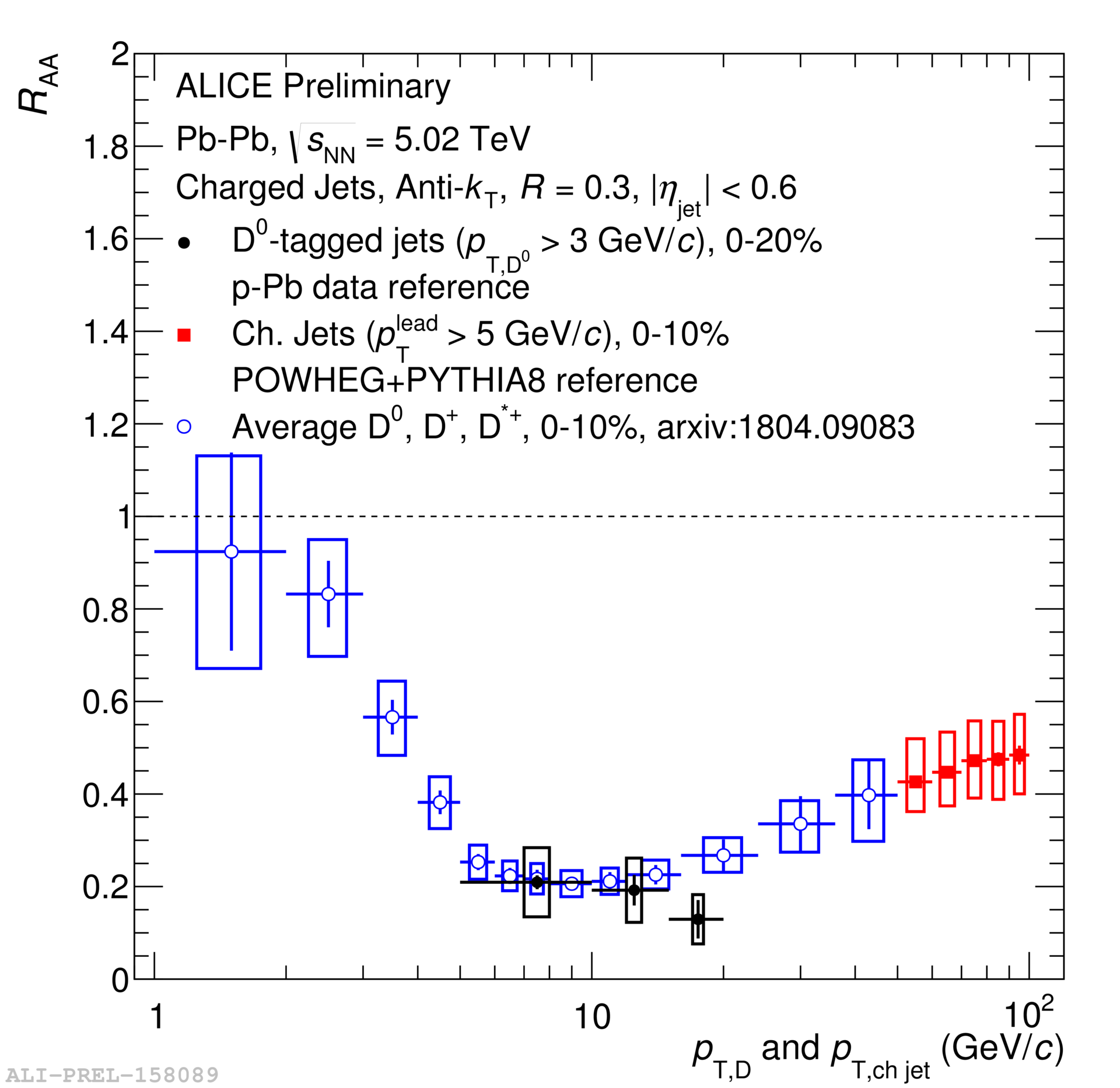}
\caption{Left: $\pt$-differential production cross section of $\Dzero$-jets compared with POWHEG+PYTHIA~\cite{Nason:2004rx,Jadach:2015mza} predictions. Right: Nuclear modification factor of $\Dzero$-jets as a function of $\pt$, compared with those of average D-mesons and inclusive jets.}
\label{Djets}       
\end{figure}

In the left panel of Fig.~\ref{Djets}, the $\Dzero$-jet production cross section in p--Pb collisions is showed for $5 < \pt({\rm jet}) < 50$ $\gevc$. Data are compared to POWHEG+PYTHIA predictions, which are found to be compatible with ALICE measurements within uncertainties, dominated by the theory uncertainty band.
In the right panel of the same figure, the nuclear modification factor of $\Dzero$-jets in 0--20\% central Pb--Pb collisions is presented for $5 < \pt({\rm jet}) < 20$ $\gevc$, showing a strong suppression by a factor $\approx$ 5-6. $\Dzero$-jet $R_{\rm AA}$ is compared with the inclusive-jet $R_{\rm AA}$, dominated by gluon and light-flavour jets. A stronger suppression seems to occur for $\Dzero$-jets, though the non-overlapping $\pt$ ranges of the two measurements prevent drawing firm conclusions. $\Dzero$-jet $R_{\rm AA}$ is also compared with the average of non-strange D-meson $R_{\rm AA}$~\cite{Acharya:2018hre}, showing a similar suppression in the common $\pt$ range. This could signal that the $\Dzero$-jets energy loss could be dominated by the energy loss of their leading particle, although the different energy scale for the charm quark transverse momentum in the two cases has to be accounted for.

\bibliography{bibliography}
%
%
%
%
%

\end{document}